\documentclass[11pt]{article}
\usepackage[margin=1in]{geometry}
\usepackage{xcolor}
\usepackage{amsmath,amssymb,latexsym,bm}
  \usepackage[T1]{fontenc}
  \usepackage[utf8]{inputenc}
  \usepackage{textcomp} 
\usepackage{booktabs,array}
\usepackage{caption}
\usepackage{graphicx}
\usepackage{bookmark}
\usepackage{hyperref,url}
\usepackage{authblk}
\usepackage[super,numbers,sort&compress]{natbib}

\DeclareMathOperator{\odiv}{\operatorname{div}}

\newcommand{\brho}{\boldsymbol{\rho}}
\newcommand{\bvarphi}{\boldsymbol{\varphi}}
\newcommand{\bpsi}{\boldsymbol{\psi}}

\title{Basis Functions for Time-Dependent Kohn-Sham Inversion}

\author[1]{Paul M. Zimmerman\thanks{Corresponding author. Email: paulzim@umich.edu}}
\author[1]{Joshua Kammeraad}
\author[1]{Duy-Khoi Dang}
\author[2]{Harish S. Bhat\thanks{Corresponding author. Email: hbhat@ucmerced.edu}}
\affil[1]{Department of Chemistry, University of Michigan, Ann Arbor, MI}
\affil[2]{Department of Applied Mathematics, University of California Merced,
Merced, CA}

\date{\today}

\begin{document}

\maketitle

\begin{abstract}
Floquet theory provides insight into the inversion of time-dependent
Kohn-Sham density functional theory. Specifically, mathematical
derivations show that the fundamental frequencies of a time-dependent
wavefunction solution are the leading-order harmonics for the TD-KS
state. Numerical tests of the resulting ansatz in 1D and 3D for atomic
and molecular cases demonstrate its utility. In particular, low
\(L_{2}\) errors in the time-dependent density and longitudinal current
were found, even though currents were not an explicit optimization
objective. In all, the proposed inversion ansatz provides
exchange-correlation potentials from time-dependent wavefunctions, is
highly interpretable, and may significantly help in the development of
nonadiabatic density functionals.
\end{abstract}

\section{Introduction}

Time-dependent Kohn-Sham (TD-KS) theory, also known as real-time
time-dependent density functional theory (DFT), describes the evolution
of electronic states of molecules and materials with an effective
balance of computational cost and accuracy. TD-KS operates as a
single-particle theory but can be just as exact as many-body TD wave
function (TD-WF)
theory.\cite{runge1984a,marques2004a,li2020a} Reaching this limit, however, requires an exact nonadiabatic\cite{maitra2002a,maitra2016a} density functional to model quantum effects of electrons, and no such functional is currently known.
Building good models is a challenge because the exact adiabatic (ground-state) functional is not available\cite{becke2014a,kanungo2025a} and even less is known about the nonadiabatic contributions to the functional.\cite{raghunathan2012a,habenicht2014a,quashie2017a,lacombe2023a}.

One potential path to better TD-KS computations is to invert TD-WFs into
the KS realm, derive relevant KS quantities, and use this information as
a guide for modeling nonadiabatic functionals.

Consistent with this goal, previous studies have inverted the TD-KS
equations.\cite{d1999a,lein2005a,ullrich2006a,thiele2008a,elliott2012a,ramsden2012a,ruggenthaler2013a,thiele2014a,ruggenthaler2015a,fuks2016a,jensen2016a,hodgson2016a,suzuki2017a,nielsen2018a,lacombe2020a,dar2022a,bhat2022a} To perform an inversion,
the time-evolving KS density \(\rho^{KS}(r,t)\) and current
\(J^{KS}(r,t)\) are derived while enforcing a match between KS and WF
densities, under a continuity condition. This can be done, for example,
by invoking a single-orbital KS state with two electrons that appears as
a product of magnitude and complex phase,
\(\phi(r,t) = \sqrt{\rho(r,t)/2}\ e^{i\alpha(r,t)}\). Here, the input
\(\rho(r,t)\) would come from a wave function calculation and
\(\alpha(r,t)\) would be solved for numerically, with
\(\rho^{KS}(r,t) = {2\left| \phi(r,t) \right|}^{2}\). Most TD-KS
inversions have been done for simple models (often in 1 or 2
dimensions), with the notable exception of a full inversion of the 3D
helium atom.\cite{dar2021a} As pointed out
by these works, there is interesting and potentially useful information
available from inverted TD-KS states. This includes steps and peaks in
the KS potential\cite{dar2022a} which mirror strong correlation
effects that are also present in inverse ground-state
DFT.\cite{kohut2016a,hodgson2017a,khanna2026a}

TD-KS shares significant similarities to TD-WF theory, with some drastic
differences. For instance, take the electronic Hamiltonian,
\(\widehat{H} = {\widehat{H}}_{0} + {\widehat{H}}_{t}\), and consider
time evolution under a fixed external potential,
\({\widehat{H}}_{t} = 0\). While \({\widehat{H}}_{0}^{WF}\) is \emph{a
priori} known and does not change with time, the situation for TD-KS
theory is more nuanced. First, \({\widehat{H}}_{0}^{KS}\) for a
superposed electronic state is unknown, and in general depends on the
initial state. This means \({\widehat{H}}_{0}^{KS}\) is not derivable
from ground-state information, though ground-state eigenstates can be
invoked to help with time
propagation.\cite{wang2015a,ma2015a,ghosal2022a} While in TD-WF theory
eigenstates are straightforwardly available from diagonalization of
\({\widehat{H}}_{0}^{WF}\), even for the exact nonadiabatic density
functional \({\widehat{H}}_{0}^{KS}\) would still be a function of
time.\cite{marques2004a} Regardless of these differences, inversion of
the TD-KS equations could lead to insight into the structure of the
exact \({\widehat{H}}_{0}^{KS}\), so bridging the gaps between TD-WF and
TD-KS theories is merited.

Herein we consider whether the structure of TD-WF solutions might help
form a basis for TD-KS inversion. While not a typical approach for TD-KS
models, TD-KS solutions might be written as a superposition of states in
the following form:
\begin{equation}
\label{eqn:ansatz}
\phi(r,t) = \sum_{n} b_{n} e^{-i \omega_{n} t} \varphi_{n}(r)
\end{equation}
where \(b_{n}\) are scalar expansion coefficients, \(\omega_{n}\) are
angular frequencies, and \(\varphi_{n}\) are the orthogonal orbitals.
This is analogous to the solutions to TD-WF problems, which would be
expressed in a many-particle basis rather than a single-particle basis
(with eigenstates \(\psi_{n}\) of \({\widehat{H}}_{0}^{WF}\) replacing
\(\varphi_{n}\)). Equation (\ref{eqn:ansatz} can be contrasted to a prescription for
evolving TD states recently introduced by Maitra and
coworkers.\cite{dar2024a} The authors suggest a basis involving
TD-DFT excited states (rather than orbitals), and show that linear
response functionals can be used to describe Rabi oscillations and
charge transfer. The success of their basis hints that the ansatz of
Equation (\ref{eqn:ansatz}) also might be of use for TD-KS theory.

This work derives and studies a basis (Equation (\ref{eqn:ansatz})) for inverse TD-KS
theory, and provides a parametrization based on TD-WF data. Results will
show that the KS basis reproduces TD-WF densities and currents to high
accuracy, measured in terms of the \(L_{n}\) errors in the density and
comparisons between current vectors. From this basis the TD
exchange-correlation potential \(V_{xc}(r,t)\) is also available, which
may inform research in modeling nonadiabatic density functionals.

\section{Theory}
\label{sect:theory}
\subsection{Time-Dependent Configuration Interaction}
For time-independent Hamiltonians, the Schr\"{o}dinger equation can be
solved in an eigenbasis,
\begin{equation}
\label{eqn:tdci}
\Psi(\mathbf{r}, t) = \sum_{n}^{}d_{n}e^{- i\omega_{n}t}\psi_{n}(\mathbf{r})
\end{equation}
where \(d_{n}\) are coefficients that specify the initial state,
\(\omega_{n}\) are angular frequencies, and \(\psi_{n}\) are
eigenvectors. The eigenvectors are expanded in a Slater determinant
basis, \(\psi_{n} = \sum_{p} {c_{p}\Phi_{p}}\) that does not depend on
time.

TD-CI provides a range of approximations to the TD-SE, from the simple
CI singles (CIS) to the highly correlated full CI (FCI). TD-FCI gives
the exact solution to the TD-SE (within its basis set) but is tractable
only for small molecules. Depending on the Hamiltonian and initial
conditions, a time-dependent problem may be qualitatively solved by CIS
(i.e., in a perturbative regime) or may require a more correlated
treatment (for strong fields).

The CIS level of theory\cite{dreuw2005a} includes singles excitations from a closed-shell initial determinant,
\[
\Psi = c_{0}\Phi_{0} + \sum_{ia}^{}{c_{ia}\Phi_{ia}}
\]
where \(i\) is an index of an occupied orbital and \(a\) of a virtual
orbital. FCI continues this expansion to all levels of electron
excitation,
\[
\Psi = c_{0}\Phi_{0} + \sum_{ia}^{}{c_{ia}\Phi_{ia}} + \sum_{i < j,a < b}^{}{c_{ijab}\Phi_{ijab}} + \cdots
\]
Whether CIS or FCI, eigenstates are found by diagonalization of the
Hamiltonian in the determinant basis. The parallel CI implementation of
ref. \citenum{dang2023a} is used herein for this purpose.

The CI structure has the flexibility needed to produce a reduced
eigenbasis and use it to (approximately) solve the TD-CI equations. For
example, one may select the lowest \(N\) eigenstates (sorted by energy)
and a physically motivated initial condition (to set \(d_{n}\)). Or, for
an initial condition that is a superposition of \(M\) arbitrary
eigenstates, only those \(M\) states are needed. In this work, TD-CI
will be treated as the reference solution to be matched by TD-KS,
irrespective of approximations made within the CI.

\subsection{Proposal for Time-Dependent KS Ansatz}
As the TD-KS equations are Schr\"odinger equations, it is possible to
represent solutions in a basis of orthogonal orbitals,
\[
\phi(r,t) = \sum_{n}^{}{b_{n}e^{-i\omega_{n}t}\varphi_{n}(r)}
\]
To illustrate how such a representation arises, we begin with an exact
derivation in the case where the TD-CI wave function is a superposition
of two eigenstates. Suppose
\({\widehat{H}}_{0}^{WF}\psi_{j}(\mathbf{r}) = E_{j}\psi_{j}(\mathbf{r})\)
and let
\(\Psi(\mathbf{r},t) = c_{0} e^{-iE_{0}t}\psi_{0}(\mathbf{r}) + c_{1}e^{-iE_{1}t}\psi_{1}(\mathbf{r})\). Let \(\rho^{WF}(r,t)\) be the corresponding 1-electron density and observe that it will be periodic with frequency
\(\Omega = E_{1} - E_{0}\) and period \(T = 2\pi/\Omega\). The analogous
TD-KS model for a two-electron system with one KS orbital would be:
\[
i\dot{\phi}(r,t) = H_{0}^{KS}\lbrack\rho\rbrack\phi(r,t) = \left( \widehat{T} + V_{H}\lbrack\rho\rbrack + V_{xc}\lbrack\rho\rbrack + V_{ext} \right)\phi(r,t).
\]
Here \(V_{H}\lbrack\rho\rbrack\), \(V_{xc}\lbrack\rho\rbrack\), and
\(V_{ext}\) are the Hartree, exchange-correlation, and external
potentials, respectively. In addition,
\(\rho = \rho^{KS} = 2|\phi|^{2}\), which feeds back into the dynamics
via the Hartree and exchange-correlation terms. Suppose that all TD-KS
terms (including \(V_{xc}\)) have been chosen such that
\(\rho^{KS}(r,t) = \rho^{WF}(r,t)\). The Runge-Gross and van Leeuwen
theorems guarantee that such an exact regime exists. In this regime
\(\rho = \rho^{WF}\) is taken as known and substituted into the TD-KS
Hamiltonian. The result is
\[
H_{0}^{E}(r,t) = H_{0}^{KS}\left\lbrack \rho^{WF}(r,t) \right\rbrack.
\]
Here \(H_{0}^{E}(r,t)\) is a Hamiltonian that depends only on space and
time. The superscript \emph{E} denotes that it is valid only in the
exact regime. By substituting in \(\rho = \rho^{WF}\), nonlinearity and
nonlocality are eliminated from the Hamiltonian. Continuing with the
choice of one doubly-occupied KS orbital, if solving
\(i\dot{\phi} = H_{0}^{KS}\lbrack\rho\rbrack\phi\) yields \(\phi\) such
that \(\rho = 2|\phi|^{2}\) matches \(\rho^{WF}\), then \(\phi\) must
satisfy the linear equation
\[
i\dot{\phi}(r,t) = H_{0}^{E}(r,t)\phi(r,t).
\]
As \(\rho^{WF}\) is time-periodic, so is \(H_{0}^{E}.\) This periodicity suggests the use of Floquet theory. Let \(U\left( t_{2},t_{1} \right)\) denote the propagator that advances the state of the linear system from time \(t_{1}\) to time \(t_{2}\). Then Floquet's theorem implies the existence of a \(T\)-periodic operator \(P\) and a Hermitian Floquet Hamiltonian \(H_{F}\) such that \(U(t,0) = P(t,0)\exp\left( - iH_{F}t \right)\) with \(P(t + T,0) = P(t,0)\) and \(P(0,0) = I\), the identity operator.\cite{shirley1965a,tannor2007a,telnov1997a} The eigenvalue problem for the Floquet Hamiltonian,
\[
H_{F}\Phi_{\alpha}(r) = \epsilon_{\alpha}\Phi_{\alpha}(r),
\]
yields both discrete and continuous spectra. For bound states, only the
discrete spectrum (or quasi-energies)
\(\left\{ \epsilon_{k} \right\}_{k \geq 0}\) and corresponding
eigenstates are important. Let \(M_{k}\) be the multiplicity of
eigenvalue \(\epsilon_{k}\). Then expand the TD-KS initial state as
\(\phi(r,0) = \sum_{k}^{}{\sum_{m = 1}^{M_{k}}c_{k,m}}\Phi_{k,m}(r)\)
with \(c_{k,m} = \left\langle \Phi_{k,m}|\phi(r,0) \right\rangle\).
Together with the Floquet decomposition of the propagator, this yields
\[\begin{aligned}
\phi(r,t) & = U(t,0)\phi(r,0) \\
 & = P(t,0)\exp\left( - iH_{F}t \right)\sum_{k}^{}{\sum_{m = 1}^{M_{k}}c_{k,m}}\Phi_{k,m}(r) \\
 & = \sum_{k}^{}{\sum_{m = 1}^{M_{k}}c_{k,m}}P(t,0)e^{- i\epsilon_{k}t}\Phi_{k,m}(r).
\end{aligned}\]
Because \(P\) is \(T\)-periodic, expand it in a Fourier series:
\(P(t,0) = \sum_{\ell}^{}e^{i\ell\Omega t}Q_{\ell}\),
where each \(Q_{\ell}\) is an operator that depends only on
space, not time. Define
\(F_{\ell,k,m} = Q_{\ell}\Phi_{k,m}.\) Then we have
\[\phi(r,t) = \sum_{\ell,k,m}^{}c_{k,m}e^{- i\epsilon_{k}t}e^{i\ell\Omega t}F_{\ell,k,m}(r).\]
Now, examining the corresponding 1-electron density,
\[
\rho(r,t) = 2\left| \phi(r,t) \right|^{2} = 2\sum_{\ell,k,m\mathcal{,l’,}k’,m’}^{}c_{k,m}c_{k’,m’}^{*}e^{- i\left( \epsilon_{k} - \epsilon_{k’} \right)t}e^{i\left( \ell - \ell' \right)\Omega t}F_{\ell,k,m}(r)F_{\ell',k’,m’}^{*}(r).
\]
As \(\rho\) must match \(\rho^{WF}\), it must be periodic with period
\(T\). The Floquet Hamiltonian's quasi-energies
\(\left\{ \epsilon_{k} \right\}\) lie in the interval
\(\lbrack 0,\Omega),\) so their differences cannot contribute integer
multiples of \(\Omega\). This means that only one quasi-energy term
(which we take to be \(e^{- i\epsilon_{0}t}\)) plays a role in the
solution \(\phi\). The KS potential is determined only up to an
arbitrary constant; without loss of generality assume this constant is
chosen such that \(\epsilon_{0} = 0\). Then
\begin{equation}
\label{eqn:phirecon}
\phi(r,t) = \sum_{\ell}^{}{\sum_{m = 1}^{M_{0}}c_{0,m}}e^{i\ell\Omega t}F_{\ell,0,m}(r)
\end{equation}
After reindexing and relabeling, we obtain precisely the representation
given in Equation (\ref{eqn:ansatz}).

Starting instead with a superposition of three states, the 1-electron
density from the exact wave function would oscillate at three
frequencies: \(\Omega_{1} = E_{1} - E_{0}\),
\(\Omega_{2} = E_{2} - E_{0}\), and \(\Omega_{3} = E_{2} - E_{1}\).
Using this density to reformulate the TD-KS Hamiltonian as \(H_{0}^{E}\)
gives a linear system that is quasiperiodic rather than periodic. In
this case, multi-mode Floquet theory is
used,\cite{sambe1973a,ho1983a,ho1984a,ho1985a,peskin1993a,jauslin1991a}
a generalization of classical Floquet theory, to
again derive the representation given in Equation (\ref{eqn:ansatz}). The critical fact
is that the propagator \(U(t,0)\) admits a decomposition into a
quasiperiodic operator \(P\) and a propagator
\(\exp\left( - iH_{F}t \right)\) associated with a Floquet Hamiltonian
\(H_{F}\). Here \(P\) can be expanded in a double Fourier series with
fundamental frequencies \(\Omega_{1}\) and \(\Omega_{2}\); such a series
suffices because \(\Omega_{3} = \Omega_{2} - \Omega_{1}\). As the number
of states comprising the superposition grows, the overall story is the
same, and the above derivation holds. To generalize to these cases,
i.e., a superposition of \(S\) states, the representation for \(\phi\)
will contain a multidimensional Fourier series over \(S - 1\)
fundamental frequencies, each of which appears in the exact wave
function and its density.

We return to Equation (\ref{eqn:ansatz}) and adopt its notation in what follows. The
structure of this representation mirrors that of the TD-CI eigenbasis,
where the spatial and temporal functions appear in product form. Because
\(V_{xc}\) is unknown at the start of our procedure, the full KS
Hamiltonian is unknown. Also, the basis orbitals \(\varphi_{n}(r)\) are
not the ground state KS orbitals, so inversion is needed to determine
coefficients \(b_{n}\) and KS eigenorbitals \(\varphi_{n}\).

By presenting orbitals, expansion weights, and characteristic
frequencies in a separable, transparent ansatz, Equation (\ref{eqn:ansatz}) reveals
hidden structure in TD-KS theory and simplifies solution of the inverse
problem. In this formulation, each frequency \(\omega_{n}\) must equal
an integer multiple of a frequency that appears in the parent TD-CI
solution. As a first approximation, unless stated otherwise, we retain
only the fundamental harmonic terms from the full Fourier series
(Equation (\ref{eqn:phirecon})). To be precise, if the parent TD-CI solution has modes at
frequencies \(\Omega_{1},\ldots,\Omega_{S}\), then our ansatz is
Equation (\ref{eqn:ansatz}) with
\(\omega_{n} \in \left\{ \Omega_{1},\ldots,\Omega_{S} \right\}.\) In a
more general setting than TD-CI, for instance where the TD-WF is not
expressed in an eigenbasis, all three quantities (\(b_{n}\),
\(\omega_{n}\) and \(\varphi_{n}\)) would need to be determined. To
parametrize (1) using TD-CI, the focus is to solve for the
time-independent eigenorbitals \(\varphi_{n}(r)\).

Equation (\ref{eqn:ansatz}) can make use of finite basis sets, which gives considerable
reduction in number of degrees of freedom compared to numerical real
space representations. If we consider \(\varphi_{n}\) to be expanded in
a basis, akin to linear combination of atomic orbitals (LCAO), the
spatial degrees of freedom can rely on existing infrastructure in
quantum chemistry programs. The use of a basis, however, means that the
TD-KS densities will not match the TD-WF densities to arbitrary
precision. This trait is the same as---and in a sense inherited
from---ground-state DFT inversion in finite basis
sets.\cite{khanna2025a} Regardless, Section \ref{sect:results} will show that high accuracy TD-KS states can nonetheless be
derived, as long as the finite basis sets are large enough.

The derivation above focuses on the dynamics of two electrons in a
single doubly-occupied KS orbital. The derivation is not intended to
apply to all possible cases: there exist systems and KS initial states
such that the Floquet-justified ansatz does not hold.\cite{kapoor2013a} This is compatible with the notion that the ansatz (\ref{eqn:ansatz}) should be justifiable for particular systems,
including those with more than two electrons. For such systems, the
computational cost of inversion via alternative techniques (e.g.,
high-dimensional constrained optimization as in ref. \citenum{bhat2026a}) will
exceed that of the procedure described in the present paper.

Having defined the KS basis, we now introduce an algorithm to solve for
\(\varphi_{n}\), built upon finite basis sets.

\subsection{Algorithm for Inversion into TD-KS Eigenbasis}

Given a TD-CI solution, the KS eigenorbitals are parametrized using
reference WF densities. The first step is to set an ansatz for the
orbitals \(\varphi_{n}(r)\). We introduce an orbital basis
\({\widetilde{\varphi}}_{m}\) and expand
\begin{equation}
\label{eqn:lcao}
\varphi_{n}(r) = \sum_{m}^{}{c_{nm}{\widetilde{\varphi}}_{m}}(r).
\end{equation}
This is deliberately analogous to LCAO. While the basis functions
\({\widetilde{\varphi}}_{m}\) could in principle be atomic orbitals, a
more compact representation is possible. The proposed procedure involves
selection of a small number of orthonormal molecular orbitals that span
the TD-KS state. This LCMO approach is not strictly required but is
computationally convenient and results in significantly reduced
computational costs.

The approach is as follows. Select \(M\) timepoints, each of which
provides a density \(\rho^{WF}(r,t).\) At each timepoint, a ground-state
inversion gives the KS orbital that accurately describes the CI density.
Orthogonalize these MOs across the \(M\) timepoints, discard any MOs
that are linearly dependent, normalize, and call the orthonormal basis
\({\widetilde{\varphi}}_{m}(r)\). Once the orbitals are available,
\(\varphi_{n}(r)\) are determined by varying \(c_{nm}\) in equation (3),
i.e., by optimizing \(\rho^{KS}(r,t)\) through ansatz (1) to match
\(\rho^{WF}(r,t)\).

To generate the \({\widetilde{\varphi}}_{m}\) orbitals in a finite
basis, the algorithm suggested by Rask et al\cite{rask2024a} is applied. This produces KS orbitals by
minimizing
\begin{equation}
\label{eqn:L}
L = \int_{}^{}{\left| \rho^{WF}(r) - \rho^{KS}(r) \right|^{2}dr\ } + \lambda T^{KS}.
\end{equation}
The densities \(\rho(r)\) are expanded over atomic orbitals, with AO
basis functions \(\chi_{\mu},\) and
\[
\rho^{KS}(r) = \sum_{\mu\nu}^{}{D_{\mu\nu}^{KS}\chi_{\mu}(r)\chi_{\nu}(r)}.
\]
The AO density matrix is given through MO coefficients \(c_{\mu i}\) as
\[
D_{\mu\nu}^{KS} = 2\sum_{i}^{}{c_{\mu i}c_{\nu i}},
\]
with the underlying AO basis expansion producing the MOs,
\[
{\widetilde{\varphi}}_{i} = \sum_{\mu}^{}{c_{\mu i}\chi_{\mu}(r)}.
\]
The basis orbitals are assumed to be real, though this is not a strict
requirement.

Finally, in Equation (\ref{eqn:L}), \(T^{KS}\) is the kinetic energy, and the
scalar \(\lambda\) controls the relative contribution of density and
kinetic energy to the definition of the KS state. In a complete basis,
the difference in density \(\rho^{WF} - \rho^{KS}\) can be optimized to
zero, and the KS state would be defined by the minimum \(T^{KS}\)
determinant that zeroes the density difference (and \(\lambda\) would
not be required). In a finite basis this is not possible, so a balance
of density minimization and kinetic energy minimization is necessary.
This algorithm has been used to provide high-accuracy KS states for
atoms helium through neon.\cite{jauslin1991a}

In all, repeated invocation of equation (4) creates a basis of KS
orbitals from input WF densities. In the context of this paper, only 1
occupied orbital is needed, even though the solver can work with more.
Further details of this KS orbital solver are available in the original
publication.\cite{khanna2025a}

Returning to the TD-KS inversion, a spatial basis is available for
Equation (\ref{eqn:lcao}) through the ground state inversions just described. Because
the basis functions \({\widetilde{\varphi}}_{m}\) obey
\(\left( {\widetilde{\varphi}}_{m} \middle| {\widetilde{\varphi}}_{n} \right) = \delta_{mn}\),
the definition of TD-KS eigenstates \(\varphi_{n}\) requires solving for
coefficients \(c_{nm}\). To maintain orthonormality, \(c_{nm}\) are
defined over orthogonal transformations
\[
c = R_{12}\left( \theta_{12} \right)R_{13}\left( \theta_{13} \right)\ldots
\]
where \(R_{ij}\left( \theta_{ij} \right)\) is a rotation matrix
connecting orbital \(i\) and \(j\). More generally, unitary rotations
could be used, though tests determined this was not necessary.

Finally, to determine \(\theta_{ij}\) and \(b_{n}\) parameters in
equation (1), minimize
\[
L_{2}^{2} = \sum_{m}^{}{\int_{}^{}{\left| \rho^{WF}\left( r,t_{m} \right) - \rho^{KS}\left( r,t_{m} \right) \right|^{2}dr\ }}
\]
where the KS densities come from
\(\rho^{KS}\left( r,t_{m} \right) = 2\left| \phi(r,t) \right|^{2}\). The
parameters \(b_{n}\) obey \(\sum_{n}^{}b_{n}^{2} = 1\) due to
normalization. This method, built upon a finite basis, will first be
compared to exact inversion in 1D, then demonstrated as useful in 3D.

\section{Computational Details}
\subsection{Computational Test of Floquet Theory Derivation via Exact
Inversion in 1D}

To validate the Floquet theory derivation presented above, a helium atom
with electrons each confined to one spatial dimension is considered. For
this model, exact numerical inversion is straightforward, enabling
quantification of the error between inverted and exact \(V_{KS}\)
potentials as a function of grid size. Here
\(V_{KS} = V_{H} + V_{xc} + V_{ext}\). Full details of the 1D inversion
are given in the Supporting Information, with a more concise overview
here.

The procedure begins by constructing a matrix Hamiltonian that operates
on wave functions defined on an equispaced \(N \times N\) grid in
two-dimensional space (one dimension per electron). The kinetic term
uses a finite-difference Laplacian matrix. All potentials are of
soft-Coulomb type: \(V(r) = \left( r^{2} + \beta^{2} \right)^{- 1/2}\)
with \(\beta^{2} = 0.3\). Diagonal matrices represent the
electron-nuclear and electron-electron potentials. The full Hamiltonian
is then a highly sparse \(N^{2} \times N^{2}\) symmetric matrix; its
first five eigenvalues/eigenvectors are computed. For
\(\Psi\left( \mathbf{r},t \right)\) as an equal superposition of the
first and fifth states, the exact 1-electron density \(\rho^{WF}\ \)and
1-electron current \(\mathbf{J}^{WF}\) are computed. Using these
quantities, a standard method from the TD-DFT literature\cite{ullrich2012a} was used to compute the exact KS potential \(V_{KS}(r,t)\) and exact initial KS
state \(\phi(r,0).\) For each grid resolution, the correctness of the
potentials was checked by solving the TD-KS equations to show
\(V_{KS}(r,t)\) and \(\phi(r,0)\) yield \(\phi(r,t)\ \)such that
\(\rho^{KS} = 2|\phi|^{2}\) matches \(\rho^{WF}\) closely.

Adding to this \(V_{KS}(r,t)\) a kinetic term (essentially a
one-dimensional second-derivative matrix) leads to the exact regime
Hamiltonian \(H_{0}^{E}\) described in the Theory section. For this
Hamiltonian, which is periodic with frequency \(\Omega = E_{5} - E_{1}\)
and period \(T = 2\pi/\Omega\), the numerical propagator over one period
is \(U(T,0) = \exp\left( - iH_{F}T \right).\) The eigenvectors of this
matrix are the Floquet states \(\Phi_{k}(r)\). From each eigenvalue
\(\lambda_{k}\) of \(U(T,0)\), the corresponding eigenvalue of \(H_{F}\)
is recovered via
\(\epsilon_{k} = - 1/T\ \text{Im}\left( {\log\lambda}_{k} \right)\). For
the 1D problem, all eigenvalues of \(U(T,0)\) have multiplicity equal to
one.

Coefficients \(c_{k} = \left\langle \Phi_{k}|\phi(r,0) \right\rangle\)
were computed, and only the coefficient with largest absolute value was
retained; this corresponds to \(k = 0\). Then, numerical solution of
\(i\dot{\phi}(r,t) = H_{0}^{E}(r,t)\phi(r,t)\) with initial condition
\(\phi(r,0) = \Phi_{0}(r)\) yields
\(\phi(r,t) = P(t,0)e^{- i\epsilon_{0}t}\Phi_{0}(r)\). Multiplying
through by \(e^{i\epsilon_{0}t}\) removes the secular term and yields
the purely periodic part \(P(t,0)\Phi_{0}(r)\). Using the FFT decomposes
this periodic part into a Fourier series, enabling construction of the
precise \(\phi(r,t)\) ansatz described in Equation
(\ref{eqn:phirecon}) of Section \ref{sect:theory}. For the purposes of testing this theory, to construct this 1D
\(\phi^{\text{ansatz}}\), we use the full Fourier expansion (up to grid
resolution). In the results section, \(\phi^{\text{ansatz}}\) is
compared against the \(\phi\) obtained from full TD-KS propagation,
along with corresponding densities and \(V_{KS}\) potentials.

\subsection{Atomic and Molecular Cases in 3D}

Reference states from TD-CI are obtained in the following fashion.
First, the zero-field Hamiltonian is diagonalized to arrive at its
ground state. At \(t \geq 0\), a static potential is added to the
Hamiltonian. With the field on, the ground state of \(t = 0\) is no
longer an eigenstate, but the field-on eigenstate basis is superposed to
specify the \(t = 0\) state as the initial state. To determine how many
eigenstates to use in the TD-CI, \(N\) states of the field-on
Hamiltonian are selected based on having the largest weights on the
coefficient vector \(d_{n}\) (c.f. Equation (\ref{eqn:tdci})). That is, the Hamiltonian
is diagonalized to obtain more than \(N\) low-lying states, then reduced
to only include terms that significantly contribute to the time
evolution. This means that states with low or no coupling under the
field are dropped. Representative initial states for the TD-CI are
delineated in the Supporting Information (Section B).

The external potentials were designed to produce a nearly uniform
electric field near the center of the atoms and molecules, which
vanishes at longer range. Specifically,
\(V(r) = a z \left( 1 + e^{(z/b)^{2}} \right)^{- 1}\), where the
coordinate system is centered at the middle of the atom or molecule. The
potential \(V(r)\) represents a near-constant electric field in the
relevant region where the electron density is high (i.e., \(z < b\)).
Its decay at long range reduces the chance that unbound states will
appear in the CI solutions, compared to a fully uniform electric field.
Herein, \(b = 5\) Bohr and \(a\) is the value of the field strength,
reported in atomic units (a.u.).

TD-CI simulations were performed using either CIS\cite{dreuw2005a} or
the select CI approach known as heat-bath
CI.\cite{holmes2016a,chien2018a} HBCI provides a convenient means for downselection of determinants from
a larger set, allowing near-FCI level accuracy with greatly reduced
computational costs. To obtain the Hamiltonian, the external potential
is evaluated in the basis set as
\(F_{\mu\nu} = \left( \chi_{\mu}(r) \right|V(r)\left| \chi_{\nu}(r) \right)\ \)using
a Becke grid,\cite{becke1988a} then transformed into
the molecular orbital basis for use in CI. Evaluation of the potential
is done via the SlaterGPU integration
library.\cite{dang2022a,stark2026a}

The TD-CI and TD-KS density currents (\(J(r,t)\)) were split into their
irrotational and divergence free components using a Helmholtz
decomposition. This was done numerically, using a finite-difference grid
to represent derivatives of the current (step size \(10^{- 8}\)). The
currents were evaluated analytically in the atomic orbital basis set.

Helium simulations (Section \ref{sect:B}) used the aug-cc-pVQZ basis
set.\cite{woon1994a} Dihydrogen simulations
(Section \ref{sect:C}) used the cc-pVQZ basis set.\cite{becke1988a}
Ethylene (Section \ref{sect:D}) was treated with a truncated cc-pVQZ basis
set, where the \(f\) atomic orbitals of hydrogen were removed, as were
the \(g\) orbitals of carbon. This resulted in a more compact basis set
for computational efficiency.

Where noted (Section \ref{sect:E}), Slater atomic orbitals were used.
Integrals to create the Hamiltonian in the Slater basis were generated
using the SlaterGPU library.\cite{holmes2016a,chien2018a} Atomic integrals
(Helium) used a grid corresponding to a product of radial functions with
a Lebedev angular grid, and molecular integrals (dihydrogen) used a
prolate spheroidal grid. Full details of the integration schemes are
available in references {[}53{]} and {[}54{]}. The Slater basis sets are
provided in the Supporting Information.

After the inversion process is completed, density errors are measured on
the same numerical grid as for the electric field evaluation (atomic
grid with Becke partitioning). Specifically,
\(L_{2} = \left( {\int_{}^{}\left| \rho^{WF}(r) - \rho^{KS}(r) \right|}^{2}dr \right)^{1/2}\)
and \(L_{1} = \int_{}^{}\left| \rho^{WF}(r) - \rho^{KS}(r) \right|dr\),
where the units are numbers of electrons. \(L_{2}\) errors in currents
are similarly evaluated, by summing over \(x,y,z\) degrees of freedom of
the \(J(r,t)\) vectors.

TD-KS states can be defined with occupancies other than that of a
closed-shell determinant.\cite{fuks2016a} Nonetheless, a
doubly-occupied KS orbital is the simplest choice and thus is used in
the results that follow.

\section{Results}
\label{sect:results}
\subsection{Computational Test of Floquet Theory Derivation via Exact
Inversion in 1D}
\label{sect:A}

Results for the helium atom in one spatial dimension are summarized in
\textbf{Table \ref{tab:helium_inversion}}. On the domain \(\lbrack - 20,\ 20\rbrack\), the grid
spacing is \(\Delta x = 40/(N - 1)\), where \(N\) is the number of grid
points in each dimension. For \(T\) as the period of superposition, the
interval \(\lbrack 0,T\rbrack\) is discretized over \(N_{T} + 1\) time
points as \(\Delta t = T/N_{T}\). For each \((\Delta x,\Delta t)\) run,
the only active eigenvalue of the Floquet Hamiltonian is
\(\epsilon_{0}\).

The \(L_{2}\) error was computed between \(\phi^{\text{ansatz}}\) and
the \(\phi^{\text{exact}}\) via propagation of the TD-KS equations, as
well as the \(L_{2}\) error between \(\rho^{WF}\) and
\(\rho = 2\left| \phi^{\text{ansatz}} \right|^{2}\).
\(\phi^{\text{ansatz}}\ \)is also used to compute
\(V_{KS}^{\text{ansatz}} = \text{Re}\left\{ \left\lbrack (1/2)\nabla^{2}\phi + i\dot{\phi} \right\rbrack/\phi \right\}\),
which is compared against the exact \(V_{KS}\). Because potentials that
differ by a constant are equivalent, at each time point, this constant
is subtracted before computing the \(L_{2}\) error in potential. For
\(\rho\), \(\phi\), and \(V_{KS}\), the \(L_{2}\) error is computed at
each of the \(N_{T}\) time points and then averaged.

\begin{table}[tbp]
    \centering
    \caption{Results from inversion of 1D helium. Definition of symbols: $\Delta x$ is grid spacing, $N_{T}$ time points, $\Delta t$ time interval, $\epsilon_{0}$ (Floquet quasi-energy), and $\| X \|_{2}$ is the $L_{2}$ error averaged over all time points.}
    \label{tab:helium_inversion}
    \setlength{\tabcolsep}{3pt}
    {\footnotesize \begin{tabular}{cccccccc}
        \toprule
        $N$ & $\Delta x$ (a.u.) & $N_{T}$ & $\Delta t$ (a.u.) & $\epsilon_{0}$ & $\| \rho^{WF} - \rho^{\text{ansatz}} \|_{2}$ & $\| \phi^{\text{ansatz}} - \phi^{\text{exact}} \|_{2}$ & $\| V_{KS}^{\text{ansatz}} - V_{KS}^{\text{exact}} \|_{2}$ \\
        \midrule
        301  & $1.33 \times 10^{-1}$ & $5 \times 10^{3}$ & $1.49 \times 10^{-3}$ & $-4.134 \times 10^{-5}$ & $1.972 \times 10^{-3}$ & $4.965 \times 10^{-3}$ & $6.039 \times 10^{-3}$ \\
        601  & $6.66 \times 10^{-2}$ & $5 \times 10^{3}$ & $1.49 \times 10^{-3}$ & $\phantom{-}2.121 \times 10^{-7}$ & $3.595 \times 10^{-4}$ & $9.058 \times 10^{-4}$ & $7.365 \times 10^{-3}$ \\
        1201 & $3.33 \times 10^{-2}$ & $5 \times 10^{4}$ & $1.49 \times 10^{-4}$ & $\phantom{-}9.627 \times 10^{-7}$ & $2.500 \times 10^{-4}$ & $5.851 \times 10^{-4}$ & $5.843 \times 10^{-4}$ \\
        2401 & $1.66 \times 10^{-2}$ & $5 \times 10^{4}$ & $1.49 \times 10^{-4}$ & $\phantom{-}2.786 \times 10^{-7}$ & $2.512 \times 10^{-5}$ & $6.532 \times 10^{-5}$ & $5.742 \times 10^{-4}$ \\
        4801 & $8.33 \times 10^{-3}$ & $2 \times 10^{5}$ & $3.72 \times 10^{-5}$ & $\phantom{-}7.402 \times 10^{-8}$ & $1.001 \times 10^{-5}$ & $2.442 \times 10^{-5}$ & $1.436 \times 10^{-4}$ \\
        \bottomrule
    \end{tabular}}
\end{table}

Overall, the results show that \(\phi^{\text{ansatz}}\), constructed
using only the single Floquet state corresponding to \(\epsilon_{0}\),
leads to a highly accurate and converging approximation to exact
densities, exact TD-KS states, and exact KS potentials. Additionally,
\(\epsilon_{0}\) itself is converging to zero, justifying the omission
of any secular \(e^{- i\epsilon_{0}t}\) term from the ansatz used in the
rest of this paper.

\subsection{The Helium Atom in 3D}
\label{sect:B}
To continue the investigation of the proposed TD-KS ansatz, TD-CI
simulations at the CIS level of theory for the helium atom in 3D were
performed. Fields of 0.15 and 0.30 a.u. along the \(z\) direction were
modeled with 3- and 5-state TD-CIS dynamics, respectively. After
shifting the CIS eigenvalues such that \(\omega_{1} = 0\), the
oscillator frequencies \(\omega\) ranged from 0 to 2 Hartree. In all,
\(\rho^{CI}(r,t)\) profiles for \(t \geq 0\) were available for
inversion to TD-KS. Details of the TD-CI simulations are given in the
Supporting Information.

From the TD-CI solutions, select time points (\(t_{m}\)) and their
respective densities were chosen. These times ranged from \(t_{m} = 0\)
to \(t_{m} = 15\) a.u. for the weaker field, and through \(t_{m} = 27\)
a.u. for the stronger field. These intervals deliberately included
sufficient time for each of the \(\omega_{n}\) periods to significantly
evolve and contribute to the span of densities. For the weaker field,
\(L_{2}\) variations in the density compared to \(\rho^{WF}(r,0)\) were
\(0.010\) on average with a peak of \(0.018\). For the stronger field,
\(L_{2}\) averaged \(0.021\) and rose as high as \(0.035\). To describe
these variations in the density, 3 orthonormal orbitals were derived
through Equation (4) for the weaker field, and 6 orbitals for the
stronger field.

Inversion of the TD-CIS resulted in good solutions to TD-KS, measured in
terms of the error in the density across the time interval. The RMS
errors were measured across the select timepoints as
\[
E_{RMS} = \left\lbrack \frac{1}{M}\sum_{m}^{}{\int_{}^{}{\left| \rho^{CI}\left( r,t_{m} \right) - \rho^{KS}\left( r,t_{m} \right) \right|^{2}dr\ }} \right\rbrack^{1/2}
\]
and are reported in \textbf{Table \ref{tab:td_summary}}. The RMS \(L_{2}\) values of order
\(10^{- 4}\) are remarkably accurate, as are the RMS \(L_{1}\) errors
near or below \(10^{- 3}\), each compared to its TD-CIS reference. For
comparison, ground-state inversions by the RKS method---which is an
accurate, finite basis set approach---reported \(L_{1}\) errors of order
\(10^{- 4}\) for the helium atom,\cite{ryabinkin2015a} similar to
the last column of \textbf{Table \ref{tab:td_summary}}.

The last row of \textbf{Table \ref{tab:td_summary}} shows an inversion of a 5-state TD-FCI,
where the CI was solved in the fully correlated limit. RMS errors are
lower than the corresponding TD-CIS case (lower field), likely because
more states were involved in the TD-CI (and therefore TD-KS solutions),
giving more degrees of freedom to fit the evolution of the density over
time.

The TD-KS solutions were tested for their ability to extrapolate to
times unseen during the inversion process. \textbf{Table \ref{tab:tdks_benchmarks}} gives RMS
\(L_{2}\) errors for TD-KS densities evaluated past the \(15\) and
\(27\) a.u. intervals used during the inversion. The errors remain quite
close to those in \textbf{Table \ref{tab:td_summary}}. The KS and CI current vectors
(\(J^{KS}(r,t)\) and \(J^{CI}(r,t)\)) are not generally the same, as
only the longitudinal component of the currents is required to
match.\cite{dar2021a} Therefore the longitudinal components of the
TD-KS and TD-CI currents were evaluated as additional benchmarks. As
\textbf{Table \ref{tab:tdks_benchmarks}} shows, the TD-KS ansatz reproduces this piece of the
currents to high accuracy, despite the current not appearing anywhere in
the optimization procedure.

\begin{figure}[tbp]
\centering\includegraphics[clip,trim=130 290 150 290,width=4in]{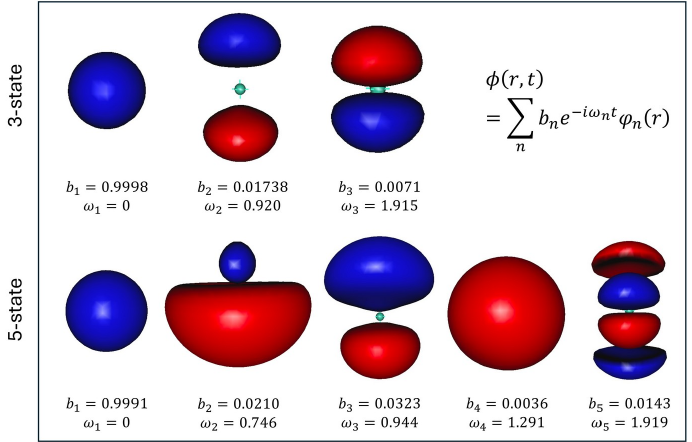}
\caption{Kohn-Sham eigenbasis (\(\varphi_{n}\)) for 3-state (upper) and 5-state (lower) dynamics of helium. Isovalue was set to 0.05. Frequencies are in atomic units.}
\label{fig:image1}
\end{figure}

The inverse TD-KS solutions are not only accurate but can be
qualitatively interpreted. Equation (\ref{eqn:ansatz}) describes a superposition of \(N\)
eigenstates, which are shown in \textbf{Figure \ref{fig:image1}} for the TD-CIS cases.
The eigenorbitals consist predominantly of \(s\) and \(p\) orbital
character, reflecting the interactions of \(s\) and \(p\) states of the
TD-CI. Comparing the 3-state TD-KS (weaker field) and 5-state (stronger
field) cases, it is interesting to see the eigenbasis sets are not
equivalent. Only the lowest energy orbitals (\(\varphi_{1}\)) are
qualitatively the same, but the others are not. While none of the TD-KS
orbitals are completely symmetric across the \(z\) axis, the asymmetry
is much more pronounced for the 5-state (stronger field) case. In
addition, the highest \(\omega\) contributors to the TD-KS states at
\(\omega \approx 2\) Ha are \(2p\) (weaker field) and \(3p\) (stronger
field) orbitals, reflecting the difference in accessible states at
different field strengths.

\begin{table}[tbp]
    \centering
    \caption{Summary of TD-CI and TD-KS for the helium atom. Max $L_{2}$ refers to the maximum difference in density ($\rho^{CI}(t) - \rho^{CI}(0)$) of all $t_{m}$ compared to the $t = 0$ CI state. RMS $L$ values compare the KS to the CI densities. All units are atomic units.}
    \label{tab:td_summary}
    \begin{tabular}{l c c c c c c c}
        \toprule
        \multicolumn{6}{c}{Reference TD-CI} & \multicolumn{2}{c}{Inverted TD-KS} \\
        \cmidrule(lr){1-6} \cmidrule(l){7-8}
        Theory & $E$-field & States & Points & Interval & Max $L_{2}$ & RMS $L_{2}$ & RMS $L_{1}$ \\
        \midrule
        CIS & 0.15 & 3 & 10 & 0 to 15.0 & $1.8 \times 10^{-2}$ & $1.9 \times 10^{-4}$ & $1.1 \times 10^{-3}$ \\
        CIS & 0.30 & 5 & 18 & 0 to 27.0 & $3.5 \times 10^{-2}$ & $1.7 \times 10^{-4}$ & $1.1 \times 10^{-3}$ \\
        FCI & 0.15 & 5 & 10 & 0 to 15.0 & $1.7 \times 10^{-2}$ & $9.8 \times 10^{-5}$ & $6.9 \times 10^{-4}$ \\
        \bottomrule
    \end{tabular}
    
    \vspace{2em}
    
    \caption{Additional benchmarks of the quality of the TD-KS states for helium. The time intervals in the 3rd column are outside of the range used in parametrizing the TD-KS state (c.f.\ Table \ref{tab:td_summary}). The final two columns are from data within the original time interval. The mean angle is the average of the angle between the vectors $J^{KS}$ and $J^{CI}$ at selected time points. All units are atomic units, except the angles in the last column.}
    \label{tab:tdks_benchmarks}
    \begin{tabular}{l c c c c c}
        \toprule
        Theory & $E$-field & Ext. Interval & Ext. RMS $L_{2}$ & Current RMS $L_{2}$ & Mean Angle \\
        \midrule
        CIS & 0.15 & 15.6 to 21.0 & $1.9 \times 10^{-4}$ & $1.0 \times 10^{-5}$ & $0.9^\circ$ \\
        CIS & 0.30 & 27.6 to 29.4 & $1.0 \times 10^{-4}$ & $6.4 \times 10^{-6}$ & $2.2^\circ$ \\
        FCI & 0.15 & 15.6 to 21.0 & $1.1 \times 10^{-4}$ & $3.1 \times 10^{-6}$ & $8.1^\circ$ \\
        \bottomrule
    \end{tabular}
\end{table}

\subsection{The Dihydrogen Molecule}
\label{sect:C}
In this example, an electric field is positioned along the bonding axis
of dihydrogen, giving time evolution of the electrons via TD-CIS. The
\(\sigma\) orbital of dihydrogen is more polarizable than the \(1s\)
orbital of helium, so only the weaker field was applied (0.15 a.u.). The
TD-CI models are expanded through 3, 5, 8, and 10 eigenstates. The
maximum \(\omega\) values for these models are respectively 0.6, 1.1,
3.4, and 7.6 Ha.

Key results for dihydrogen inversion are presented in \textbf{Table \ref{tab:dihydrogen_summary}}.
As with the helium inversions, discrete timepoints were selected to
optimize the TD-KS states (see Supporting Information for specifics).
Along these points, the CI density varied by around \(10^{- 2}\) in the
\(L_{2}\) metric. The RMS \(L_{2}\) errors for the TD-KS solution show
accuracies better than \(10^{- 4}\) can be achieved, with similarly low
RMS \(L_{1}\) errors as well. Overall, no additional difficulty was
noticed for inversion of a diatomic in 3 dimensions.

\begin{figure}[tbp]
\centering
\includegraphics[clip,trim=50 100 100 100,width=4in]{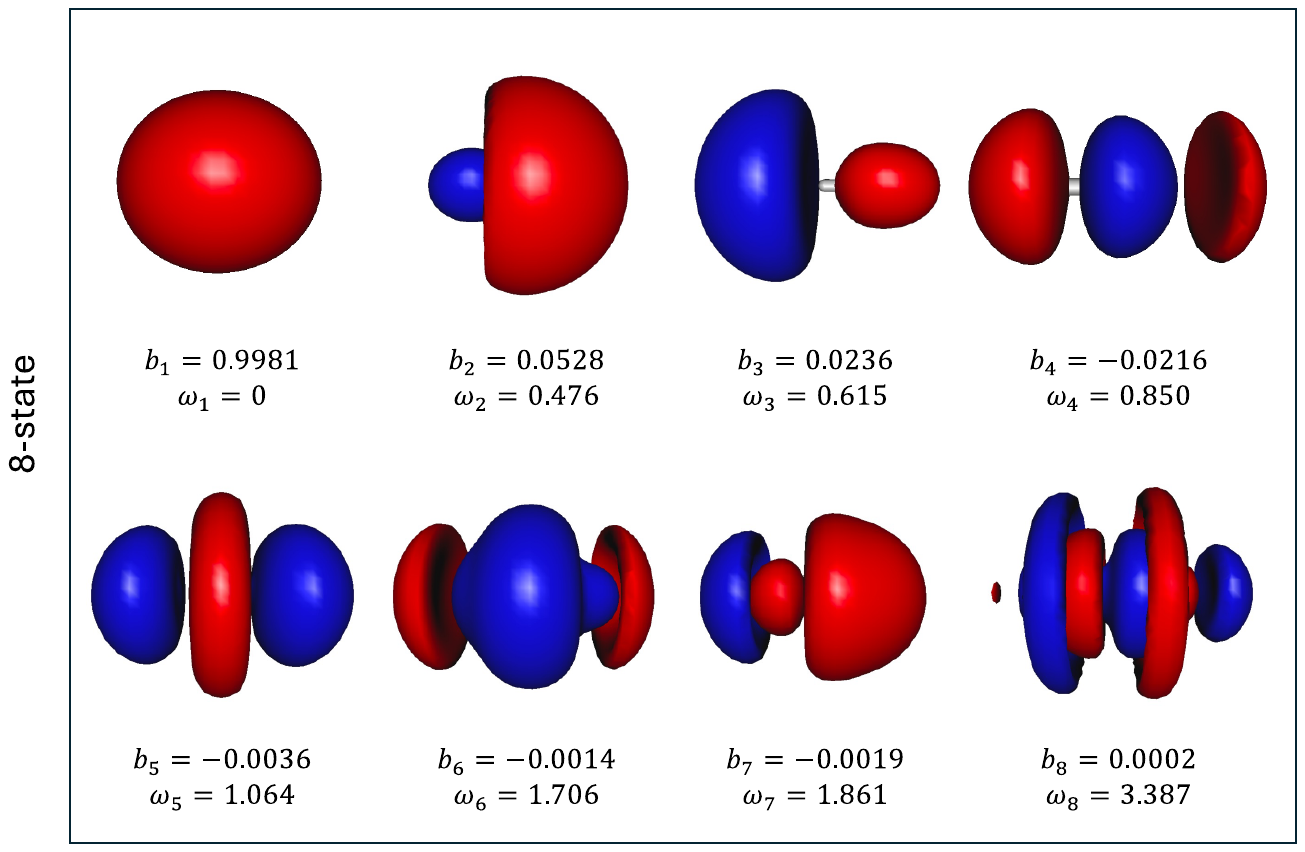}
\caption{Kohn-Sham eigenbasis (\(\varphi_{n}\)) for 8-state
dynamics of dihydrogen. Isovalues are 0.05, except for \(\varphi_{6}\),
which is isovalue 0.03. Oscillator frequencies are in atomic units.}
\label{fig:image2}
\end{figure}

Figure \ref{fig:image2} shows the orbital eigenbasis underlying the 8-state solution
for \(\phi(r,t)\). The orbitals are symmetric around the cylindrical
rotation axis, and have a variety of nodal structures along the z
direction. The \(b_{n}\) expansion coefficients show that the 4 highest
\(\omega_{n}\) basis states contribute less than the other 4 states,
since \(\left| b_{n} \right| < 4 \times 10^{- 3}\) for \(n > 4\). This
resembles the TD-CI solution, where the 4 states with the higher
\(\omega_{n}\) values have
\(\left| d_{n} \right| < 6 \times 10^{- 3}\).

The above analysis inspired this question: could the TD-KS state be
expanded in a smaller number of eigenbasis functions compared to the
TD-CI? To answer this, the 8-state TD-CI was inverted into a 4-state
TD-KS basis, choosing the 4 lower energy states only. The RMS \(L_{2}\)
errors increased to \(1.2 \times 10^{- 3}\), indicating reduction in
inversion accuracy of the same order of magnitude as the \(b_{n}\)
coefficients that were dropped. Interestingly, the 4 eigenorbitals were
essentially the same as their counterparts from 8-state inversion.
Quantitatively, the angles between the orbitals (computed via their
overlaps as
\(\cos^{- 1}\left( \varphi_{n}^{N = 8}|\varphi_{n}^{N = 4} \right)\))
were 0.02, 0.80, 1.11, and 0.83 degrees for \(n\) ranging from 1 to 4.
This finding suggests that TD-KS solutions in a reduced eigenbasis can
be good approximations to the full TD-CI dynamics.

\begin{table}[htbp]
    \centering
    \caption{Summary of TD-CI and TS-KS results for dihydrogen. Reference results from TD-CIS. Max $L_{2}$ refers to the maximum difference $\rho^{CI}(t) - \rho^{CI}(0)$. Mean angle is the average angle between the vectors $J^{KS}$ and $J^{CI}$.}
    \label{tab:dihydrogen_summary}
    \begin{tabular}{l c c c c c c}
        \toprule
        & \multicolumn{3}{c}{Reference TD-CI} & \multicolumn{3}{c}{Inverted TD-KS} \\
        \cmidrule(lr){2-4} \cmidrule(l){5-7}
        $E$-field & States & Points & Max $L_{2}$ & RMS $L_{2}$ & RMS $L_{1}$ & Mean Angle \\
        \midrule
        0.15 & 3  & 22 & $3.1 \times 10^{-2}$ & $3.2 \times 10^{-5}$ & $3.4 \times 10^{-4}$ & $0.4^\circ$ \\
        0.15 & 5  & 22 & $4.5 \times 10^{-2}$ & $7.4 \times 10^{-5}$ & $5.9 \times 10^{-4}$ & $2.4^\circ$ \\
        0.15 & 8  & 22 & $4.6 \times 10^{-2}$ & $6.5 \times 10^{-5}$ & $7.0 \times 10^{-4}$ & $1.0^\circ$ \\
        0.15 & 10 & 22 & $4.7 \times 10^{-2}$ & $6.7 \times 10^{-5}$ & $6.8 \times 10^{-4}$ & $2.8^\circ$ \\
        \bottomrule
    \end{tabular}
\end{table}

\subsection{Ethylene HOMO Evolution}
\label{sect:D}
To give insight into polyatomic TD-KS states, TD-CIS was performed for
the ethylene molecule (C\textsubscript{2}H\textsubscript{4}). The CI
treated 14 electrons as fixed within their Hartree-Fock orbitals, and
the remaining 2 electrons were evolved in time under static fields. The
active electrons experienced the full Hartree and exchange potentials of
the core electrons, and were kept strictly orthogonal to the core to
maintain Pauli exclusion. TD-KS inversion was otherwise performed as
before. Since the HOMO of ethylene is more polarizable than the
remaining orbitals, the frozen orbital model is expected to
qualitatively resemble the full, unfrozen case. The results are
presented keeping these model limitations in mind.

5-state TD-CIS solutions for ethylene were inverted to produce viable
TD-KS states with 5 KS orbitals. Three inversions were performed, one
each for fields in the \(x\), \(y\), and \(z\) directions. RMS \(L_{2}\)
errors of order \(10^{- 4}\) over a time interval of 20 a.u. indicate
little additional difficulty in treating TD-KS states in a polyatomic,
compared to atoms and diatomics.

\begin{figure}[tbp]
\centering
\includegraphics[clip,trim=140 160 140 140,width=4in]{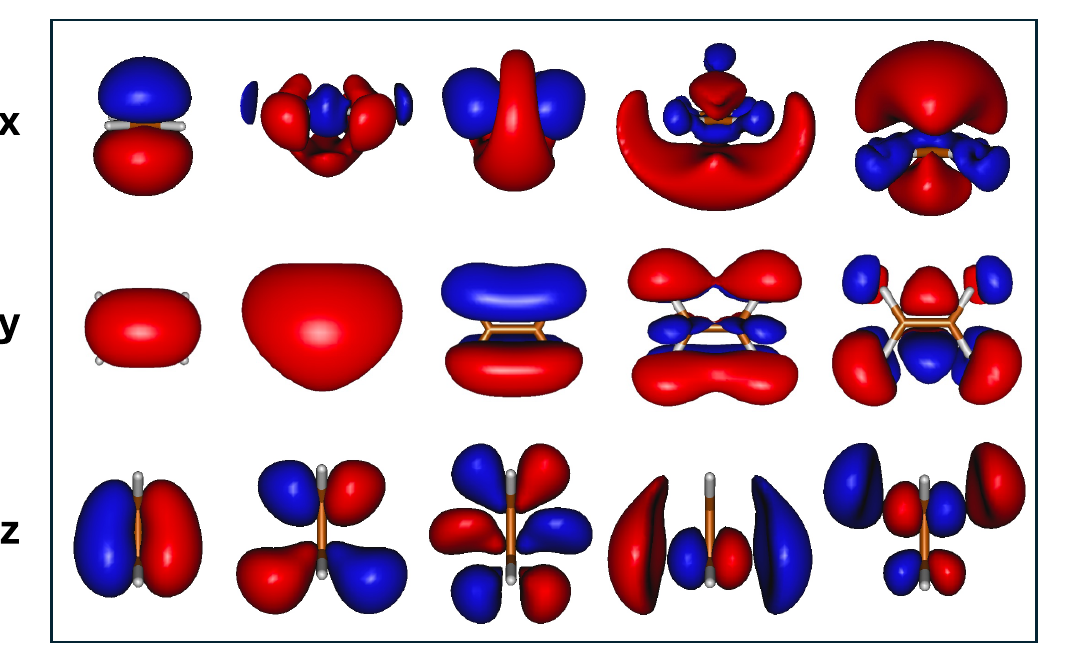}
\caption{Ethylene HOMO evolution. Molecules were oriented
uniformly within each row so the field aligns along the vertical axis.
Note that while not visible, the \(y\)-field orbitals all have a single
node in the \(yz\) plane. Isovalue was set to 0.05 for \(y\) and \(z\)
fields, and 0.03 for the \(x\) field.}
\label{fig:image3}
\end{figure}

Figure \ref{fig:image3} presents the shapes of the 15 TD-KS basis states across the 3
different fields. As seen in the figure, the possibilities for evolution
of a single $\pi$ orbital are numerous. None of the orbitals correspond to
any KS ground-state eigenstate, since they all break symmetry from the
\(D_{2h}\) point group of the molecule. While all the eigenbasis sets
contain one orbital that qualitatively matches the ground-state HOMO,
the remaining orbitals introduce polarity in the three Cartesian
directions, as appropriate to the applied field.

\subsection{Time-Dependent Exchange Correlation Potentials}
\label{sect:E}
The TD-KS equations can be directly inverted into KS potentials via
\[
v_{KS}(r,t) = \rho(r,t)^{- 1}\left\lbrack \phi^{*}(r,t)\nabla^{2}\phi(r,t) + 2i\phi^{*}(r,t)\dot{\phi}(r,t) \right\rbrack
\]
using the solutions from Equation (\ref{eqn:ansatz}). However, this procedure will not
produce accurate potentials in Gaussian basis sets, as the Laplacian is
poorly behaved.\cite{khanna2026a} For this subsection, Slater atomic
orbitals---which give substantially better properties in finite basis
set inversions than Gaussians\cite{jauslin1991a}---was used instead.
The \(V_{KS}(r,t)\) were then transformed into \(V_{xc}(r,t)\) by
subtracting out the external and Hartree terms:
\[
V_{XC}(r,t) = V_{KS}(r,t) - V_{H}(r,t) - V_{ext}(r,t)
\]
where \(V_{ext}(r,t)\) included the nuclear potential and the field
potential.

\begin{figure}[tbp]
\centering
\includegraphics[clip,trim=130 300 150 300,width=6in]{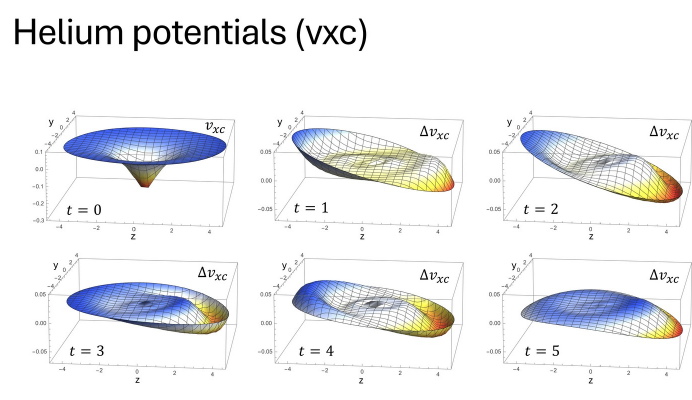}
\caption{Helium atom \(V_{xc}(r,t)\ \) and
\(\Delta V_{xc}(r,t) = V_{xc}(r,t) - V_{xc}(r,0)\) in the \(yz\) plane,
obtained by inverting \(\phi(r,t)\) from a 5-state TD-CIS model. Field
is in the \(z\) direction with magnitude 0.1 a.u. All units are atomic
units.}
\label{fig:image4}
\end{figure}

\textbf{Figure \ref{fig:image4}} shows snapshots of \(V_{xc}(r,t)\ \)for the helium
atom under a field of 0.1 a.u. from a 5-state TD-CIS reference. The
first frame contains the potential at \(t = 0\), and the others show the
deviation in the potential from this timepoint in steps of
\(\Delta t = 1\) a.u. Overall, considerable changes to the potential
occur in the first 5 time steps, reflecting the superposition of 5
orbitals in the TD-KS solution. Because \(V_{xc}\) is expressed as a
linear combination of periodic temporal functions each multiplied by
spatial basis functions, it is smooth in space and periodic in time,
avoiding numerical noise that can be problematic for inversion methods
on spatial grids. The plotted \(V_{xc}\) is relatively stable near the
nucleus, since relative changes in density are small in that region. The
longer-range features (\(r > 2\)) reflect changes in the denominator of
the inverse KS equation, i.e., \(\rho(r,t)\), where small changes in
small densities are amplified. This is a consequence of \(s\) and \(p\)
waves interfering with one another in Equation (\ref{eqn:ansatz}). Nonetheless, the
long-range part of \(V_{xc}\) would be challenging to describe with a
conventional density functional, built on semilocal features, just as it
is for ground-state DFT.

\begin{figure}[tbp]
\centering
\includegraphics[clip,trim=110 310 130 300,width=6in]{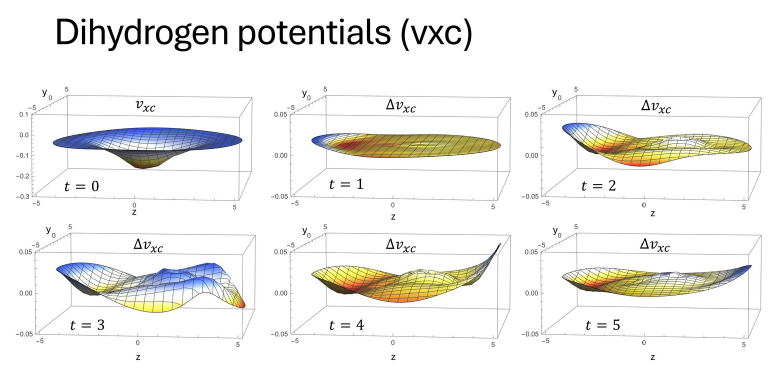}
\caption{Dihydrogen atom \(V_{xc}(r,t)\) and
\(\Delta V_{xc}(r,t)\) in the \(yz\) plane, obtained by inverting
\(\phi(r,t)\) from a 5-state TD-CIS model. The field is in the \(z\)
direction with magnitude 0.1 a.u. All units are atomic units.}
\label{fig:image5}
\end{figure}

\textbf{Figure \ref{fig:image5}} shows \(V_{xc}(r,t)\ \)for dihydrogen for the same
time range as \textbf{Figure \ref{fig:image4}}, based on inversion of TD-CIS in a
5-state basis. Once again, our approach yields \(V_{xc}\) functions with
reasonable smoothness in space and time. This is particularly remarkable
in low-density regions where the inversion problem is ill-posed. In
effect, the proposed inversion framework implicitly regularizes the
computed \(V_{xc}\), removing the need for explicit regularization terms
(e.g., penalties on \(\left\| \nabla V_{xc} \right\|^{2}\)) in the
inversion framework.

\section{Discussion}
The proposed ansatz for TD-KS inversion may be surprising for its
combination of simplicity and effectiveness. When building a basis set,
separation of variables is a common choice, though this assumption would
be more obvious for a time-independent Hamiltonian than for
\({\widehat{H}}^{KS}\). Regardless, the Floquet derivation provides a
strong theoretical justification for separation of variables, and
numerical validations are consistent with this mathematical insight
(c.f. Section \ref{sect:A}). While Floquet theory predicts that any
positive integer multiples of the TD-CI oscillator frequencies can
contribute to the TD-KS state, at least for the examples of this work,
only 1 integer was enough.

The ansatz provides some degree of protection against numerical
instabilities in the KS inversion process. There are no sharp changes in
the KS state over time, and low density regions do not unduly influence
the optimization. Regularization beyond the assumption of a finite
number of oscillator frequencies is not required, which is intrinsic to
the Floquet derivation nonetheless. Remarkably, the longitudinal
currents from the inverse solutions match those of the parent TD-CI,
even though currents were not invoked in the optimization strategy. This
contrasts with prior efforts to invert TD-KS equations that made use of
continuity equations as their underlying principle.

In addition, Equation (\ref{eqn:ansatz}) extrapolates beyond the time interval of its
optimization, with the same accuracy as within the interval (c.f.
\textbf{Tables \ref{tab:td_summary}} and \textbf{\ref{tab:tdks_benchmarks}}). This provides strong evidence that
the ansatz captures the dominant contributions to the KS state.
Analogously, the optimization can make use of existing infrastructure
for KS orbital generation in commonly available, finite basis
sets.\cite{jauslin1991a,khanna2025a} This is a substantial advantage in terms of
time to solution.

Finally, the proposed ansatz provides an interpretable solution to the
TD-KS inversion problem. As seen in \textbf{Figures \ref{fig:image1}} through
\textbf{\ref{fig:image3}}, electric fields applied to atomic and molecular cases
produce polarization responses in the KS orbital states. These orbitals
mix over time, where their superpositions provide the time-evolved
density. This interpretation is close to that of TD-WF time evolution,
but simplified from a many-body into a single-particle basis.

\section{Conclusions}

Inversion of TD-KS electronic states is shown to be possible in an
ansatz based on Floquet theory. This in turn reveals an underlying
structure to the TD-KS problem, where the lowest-order Floquet energies
match the TD-CI eigenenergies. The separation of spatial and time basis
follows, significantly simplifying the optimization procedure for
inverse TD-KS theory.

The astute reader has noticed that this ansatz is particularly merited
for inversion under time-independent external fields, and---while not
demonstrated here---time-periodic external fields. Can the \(V_{XC}\)
from this inversion setting be useful in the broader case of
time-varying external fields? Fortunately, as pointed out by Maitra and
coworkers,\cite{dar2021a} these \(V_{XC}(r,t)\) are more generally
applicable. The reader is encouraged to read \emph{Appendix A} of
reference 29 for the origin of this principle, which is derived from
exact conditions of the time-dependent XC potential.

We conclude by reiterating that the proposed TD-KS ansatz can provide
insight to help build nonadiabatic density functionals, specifically by
modeling of \(V_{XC}(r,t)\) data. This work will continue as part of our
QUEST (QUantum dynamics -- moving Electrons through Space and Time)
collaboration.

\section*{Acknowledgements}
We thank Neepa Maitra, Christine Isborn, Bikash Kanungo, and Vikram
Gavini for helpful discussions. This work was supported by the U.S.
Department of Energy, Office of Science, Office of Advanced Scientific
Computing Research and Office of Basic Energy Sciences, Scientific
Discovery through Advanced Computing (SciDAC) program under Award Number
DE-SC0026088 (all theory and computation except for 1D exact inversion).
Specifically for the development of 1D exact inversion, H.S.B.
acknowledges support from the Office of Naval Research, Grant Number
W911NF-23-1-0153. This research used resources of the National Energy
Research Scientific Computing Center (NERSC), a U.S. Department of
Energy Office of Science User Facility located at Lawrence Berkeley
National Laboratory, operated under Contract No. DE-AC02-05CH11231 using
NERSC award BES-m5214. We also acknowledge computational time on the
Pinnacles cluster, including CENVAL-ARC GPU nodes, at UC Merced,
supported by NSF awards OAC-2019144 and OAC-2346744.

\section*{Supporting Information}
\renewcommand{\thesubsection}{\Alph{subsection}}
\subsection{Methodological Details for Exact Inversion}
The inversion procedure begins with the solution of the Schr\"odinger equation for a Helium atom with both electrons considered in one spatial dimension, and with the nucleus held fixed in the Born-Oppenheimer approximation.   Given particles at locations $z_1$ and $z_2$ with corresponding charges $q_1$ and $q_2$, the potential is
\[
q_1 q_2 V(z_1 - z_2),
\]
where in 1D we use the soft Coulomb potential
\begin{equation}
\label{eqn:softcoulomb}
V(r) = ( r^2 + \alpha^2)^{-1/2}.
\end{equation}
Here $\alpha \neq 0$ is a parameter that renders $V(r)$ smooth at the origin with value $V(0) = 1/|\alpha|$; the true Coulomb potential's long-range behavior of $V(r) \sim 1/|r|$ is preserved.  For $j = 1, 2$, let $x_j$ denote the coordinate of electron $j$.  Let $x_H$ denote the position of the nucleus. Then the electronic Hamiltonian (in atomic units) is
\begin{equation}
\label{eqn:ham2D}
\widehat{H} = -\frac{1}{2} \left( \frac{\partial^2}{\partial x_1^2} +  \frac{\partial^2}{\partial x_2^2} \right) 
- 2 V(x_H - x_1) - 2 V(x_H - x_2) + V(x_1 - x_2). 
\end{equation}
In all numerical results that follow, we use $\alpha^2 = 0.3$ and $x_H = 0$.  

\paragraph{Discretization.} Our goal is to compute real-space solutions of the time-independent Schr\"odinger equation
\begin{equation}
\label{eqn:tise}
\widehat{H} \psi = E \psi.
\end{equation}
We begin by specifying our computational domain, $-L \leq x_1, x_2 \leq L$,
with grid points given by $(-L + l \Delta x, -L + m \Delta x)$.  Here $\Delta x = 2L / (N-1)$.  In this work, we choose $L = 20$ and values of $N$ given in Table 1 of the manuscript.
There are a total of $N^2$ grid points: $0 \leq l, m \leq N-1$.  The continuous wave function $\psi$ is thus approximated by a complex vector $\bpsi$ of length $N^2$.

To discretize the Hamiltonian (\ref{eqn:ham2D}) on this grid, we must discretize the Laplacian. Let $D_1$ be the following finite difference matrix:
\begin{equation}
\label{eqn:findiff1d}
D_1 = \frac{1}{12 \Delta x}
\begin{bmatrix}
0  & \cdots  & \cdots  & \cdots  & \cdots  & \cdots  & \cdots & \cdots & 0 \\[2pt]
-3 & -10 & 18 & -6 & 1  &        &        &        &   \\[2pt]
1  & -8  & 0  & 8  & -1 &        &        &        &   \\
   & 1   & -8 & 0  & 8  & -1     &        &        &   \\
   &     & 1  & -8 & 0  & 8      & -1     &        &   \\
   &     &    & \ddots & \ddots & \ddots & \ddots & \ddots &   \\
   &     &    &      & 1 & -8     & 0      & 8      & -1 \\[2pt]
   &     &    &      & -1 & 6     & -18    & 10     & 3 \\[2pt]
0  & \cdots  & \cdots  & \cdots  & \cdots  & \cdots  & \cdots & \cdots & 0 
\end{bmatrix}.
\end{equation}
This is a 4th-order discretization of the one-dimensional first derivative with Neumann (zero-flux) boundary conditions.  We set
\begin{equation}
\label{eqn:D2def}
D_2 = -D_1^T D_1.
\end{equation}
Note that (\ref{eqn:D2def}) guarantees that $D_2$ will be symmetric negative semidefinite, in correspondence with its continuum counterpart, the one-dimensional $d^2 / dx^2$ operator. In the interior of the domain, $D_2$ retains 4th-order accuracy. Let $I$ be the $N \times N$ identity matrix; let $\otimes$ denote the Kronecker product.  Then we can discretize the two-dimensional Laplacian in (\ref{eqn:ham2D}) via
\[
\mathbf{D}_2 = D_2 \otimes I + I \otimes D_2.
\]
Again, $\mathbf{D}_2$ and the continuum Laplacian are both symmetric negative semidefinite. Because of the Kronecker product, $\mathbf{D}_2$ is of size $N^2 \times N^2$. We use it to form the $N^2 \times N^2$ matrix Hamiltonian
\[
\mathbf{H} = -\frac{1}{2} \mathbf{D}_2 + \mathbf{V},
\]
where $\mathbf{V}$ is a purely diagonal matrix computed by evaluating the electron-nuclear and electron-electron potentials at each of the $N^2$ grid points described above.  Note that $\mathbf{H}$ is symmetric and highly sparse; 

We compute the first $5$ eigenvector-eigenvalue pairs of $\mathbf{H}$.  Let us denote the $j$-th eigenvector as $| j \rangle$ and the $j$-th eigenvalue as $E_j$, with $j$ starting at $0$.  

The computational domain was chosen sufficiently large such that the computed eigenstates $|j \rangle$ are negligible (absolute values less than $10^{-12}$) at and near the boundary of the domain.  Grid refinement studies for the above eigenproblem show that, in particular, the eigenvalues $E_0$ and $E_4$ exhibit 4th-order convergence.

\paragraph{Time-Dependent Quantities.}  Using the eigenstates and energies computed above, we form the time-dependent superposition
\begin{equation}
\label{eqn:exactpsi}
\boldsymbol{\psi}(t) = \frac{1}{\sqrt{2}} ( e^{-i E_0 t} | 0 \rangle + e^{-i E_4 t} | 4 \rangle )
\end{equation}
From this wave function, we compute three integrated quantities.  The first is the $1$-electron density,
\begin{equation}
\label{eqn:rhodef}
\rho(x,t) = 2 \int | \Psi(x,x_2,t) |^2 \, dx_2.
\end{equation}
From this we can derive the time-derivative of the $1$-electron density:
\begin{equation}
\label{eqn:rhodotdef}
\partial_t \rho(x,t) = -2 \operatorname{Im} \int \Psi^\ast \partial_{x}^2 \Psi \, dx_2.
\end{equation}
The third is the $1$-electron current,
\begin{equation}
\label{eqn:curdef}
\mathrm{j}(x,t) = 2 \operatorname{Im} \int \Psi^\ast \partial_{x} \Psi \, dx_2.
\end{equation}
For each choice of $\Delta t$ given in Table 1 of the manuscript, 
Discretizing $x$ using the one-dimensional version of the grid defined above (with grid points $x_{j} = -L + j \Delta x$), and discretizing time with grid points $t_k = k \Delta t$ (with $k = 0, \ldots, N_T$), we compute and save to disk $\rho$, $\partial_t \rho$, and $\mathrm{j}$.   As before, $\Delta x = 2L/(N-1)$; the values of $N$, $N_T$, and $\Delta t$ are given in Table 1 of the manuscript. 

\paragraph{Exact Inversion.}   For a 2-electron system modeled by a single KS orbital, we have the time-dependent Kohn-Sham (TDKS) equation 
\begin{equation}
\label{eqn:tdksvs}
i \partial_t \varphi = -\frac{1}{2} \nabla^2 \varphi + v_s \varphi.
\end{equation}
Here $v_s = v^\text{ext} + v^\text{H} + v^\text{X} + v^\text{C}$, i.e., $v_s$ is the sum of all potentials.  Note that $v_s$ is itself shorthand for $v_s[\rho](\mathbf{x},t)$.  Let us employ 
\begin{equation}
\label{eqn:varphipolar}
\varphi = \sqrt{ \frac{ \rho }{ 2 } } e^{i \zeta}.
\end{equation}
This is the polar decomposition of $\varphi$; note that $\varphi$, $\rho$ and $\zeta$ all depend on $\mathbf{x}$ and $t$.  Our task is to substitute (\ref{eqn:varphipolar}) into (\ref{eqn:tdksvs}).  Taking a time-derivative of (\ref{eqn:varphipolar}), we obtain
\[
\partial_t \varphi = \frac{1}{2 \sqrt{2}} \rho^{-1/2} (\partial_t \rho) e^{i \zeta} + \sqrt{ \frac{ \rho }{ 2 } } e^{i \zeta} i (\partial_t \zeta).
\]
We need $\nabla^2 \varphi$; we start with the gradient:
\[
\nabla \varphi = \frac{1}{2 \sqrt{2}} \rho^{-1/2} (\nabla \rho) e^{i \zeta} + \sqrt{ \frac{ \rho }{ 2 } } e^{i \zeta} i (\nabla \zeta).
\]
We then take the divergence of both sides to obtain
\[
\nabla^2 \varphi = -\frac{\rho^{-3/2}}{4 \sqrt{2}} \| \nabla \rho \|^2 e^{i \zeta} + \frac{\rho^{-1/2}}{2 \sqrt{2}} (\nabla^2 \rho) e^{i \zeta} + \frac{\rho^{-1/2}}{\sqrt{2}} e^{i \zeta} i (\nabla \rho) \cdot (\nabla \zeta) 
 - \sqrt{ \frac{\rho}{2} } e^{i \zeta} \| \nabla \zeta \|^2 + \sqrt{ \frac{\rho}{2} } e^{i \zeta} i \nabla^2 \zeta.
\]
Putting our results for $\partial_t \varphi$ and $\nabla^2 \varphi$ into (\ref{eqn:tdksvs}), we obtain
\[
\frac{i}{2} \rho^{-1/2} \partial_t \rho - \sqrt{\rho} \partial_t \zeta = \frac{1}{8} \rho^{-3/2} \| \nabla \rho \|^2 - \frac{1}{4} \rho^{-1/2} \nabla^2 \rho  - \frac{i}{2} \rho^{-1/2} \nabla \rho \cdot \nabla \zeta + \frac{1}{2} \sqrt{ \rho } \| \nabla \zeta \|^2 - \frac{i}{2} \sqrt{\rho} \nabla^2 \zeta + v_s \sqrt{\rho}.
\]
The real part of this equation yields
\begin{equation}
\label{eqn:vssol}
v_s = - \frac{1}{8} \rho^{-2} \|\nabla \rho\|^2 + \frac{1}{4} \rho^{-1} \nabla^2 \rho -\partial_t \zeta - \frac{1}{2} \| \nabla \zeta \|^2,
\end{equation}
while the imaginary part yields
\[
\partial_t \rho = -\nabla \rho \cdot \nabla \zeta - \rho \nabla^2 \zeta.
\]
We rewrite this as
\begin{equation}
\label{eqn:zetapde}
\odiv( \rho \nabla \zeta) = -\partial_t \rho.
\end{equation}
The above derivation can be found in Appendix E of C. A. Ullrich [2012], \emph{Time-Dependent Density Functional Theory}, Oxford University Press; we have included it here to keep the document self-contained.

\paragraph{Specializing to 1D.} In one spatial dimension, the expressions simplify.  The PDE (\ref{eqn:zetapde}) reduces to
\[
\partial_x (\rho \partial_x \zeta) = -\partial_t \rho.
\]
From our expressions (\ref{eqn:rhodotdef}) and (\ref{eqn:curdef}), we can see that the continuity equation $\partial_t \rho = - \partial_x \mathrm{j}$ is satisfied.  Using this continuity equation, we can write
\[
\partial_x (\rho \partial_x \zeta) = \partial_x \mathrm{j}.
\]
Integrating both sides with respect to $x$, we have, up to an arbitrary constant,
\[
\rho \partial_x \zeta = \mathrm{j}.
\]
Using this, we can solve for $\zeta$:
\begin{equation}
\label{eqn:zeta}
\zeta(x,t) = \int_{a}^{x} \frac{\mathrm{j}(x',t)}{\rho(x',t)} \, dx'.
\end{equation}
Here $a$ is some point in the domain at which we pin the phase to be zero, i.e., $\zeta(a,t) = 0$.  The precise value of $a$ is unimportant as constant phase shifts in (\ref{eqn:varphipolar}) do not affect TDKS dynamics---in our work, we set $a = -L$, the left endpoint of the domain. To compute (\ref{eqn:vssol}), we also need the time-derivative of $\zeta$:
\begin{equation}
\label{eqn:zetadot}
\partial_t \zeta(x,t) = \int_a^x \left[ \frac{ \partial_t \mathrm{j}(x',t) }{\rho(x',t)} - \frac{ \mathrm{j}(x',t) \partial_t \rho(x',t) }{ \rho(x',t)^2 } \, \right] dx'.
\end{equation}
Here we use our saved results for $\rho$, $\partial_t \rho$, and $\mathrm{j}$---the computation of these quantities was described above.  To compute $\partial_t \mathrm{j}$, we use a simple finite-difference derivative. For the quadrature in both (\ref{eqn:zeta}) and (\ref{eqn:zetadot}), we use the cumulative trapezoidal rule.  Once we have these pieces in place, we compute (\ref{eqn:vssol}) via
\begin{equation}
\label{eqn:vs1d}
v_s = -\frac{ (\partial_x \rho)^2 }{ 8 \rho^{2} } + \frac{ \partial_{xx} \rho }{ 4 \rho } - \partial_t \zeta - \frac{ (\partial_x \zeta)^2 }{2}.
\end{equation}
As we have a two-electron system, we use exact exchange exchange: $v^\text{X} = -v^\text{H}/2$.   The Hartree potential is
\[
v^\text{H}[\rho](x) = \int V(x - x') \rho(x') \, dx',
\]
where $V$ is the soft Coulomb potential defined by (\ref{eqn:softcoulomb}).  The external potential is $v^\text{ext}(x) = -2 V(x)$.  With these ingredients, we can compute
\begin{equation}
\label{eqn:vc1d}
v^\text{C} = v_s - v^\text{ext} - \frac{1}{2} v^\text{H}.
\end{equation}
We compute and save both $v_s$ and $v^\text{C}$ on the same equispaced grids (in space and time) described above.

\paragraph{1D TDKS Propagation.} When we solve for $\zeta$ above, we gain access to $\zeta(x,0)$.  Together with $\rho(x,0)$ computed from the wave function (\ref{eqn:exactpsi}) at $t=0$, this fully determines the TDKS initial state $\varphi(x,0)$.  With this initial state, and with the correlation potential $v^\text{C}(x,t)$ that we have computed at each point in space and time, we solve 
\begin{equation}
\label{eqn:tdks1d}
i \partial_t \varphi = \left( -\frac{1}{2} \partial_x^2 + v^\text{ext} + \frac{1}{2} v^\text{H}[\rho] + v^\text{C} \right) \varphi
\end{equation}
with the Hartree and external potentials as defined above.  This propagation enables us to test the quality of the $v^\text{C}$ obtained via exact inversion.

To solve (\ref{eqn:tdks1d}) on the one-dimensional equispaced grid defined above, we approximate the kinetic operator by $-(1/2) D_2$. To compute the Hartree potential on a discrete grid, we evaluate the kernel $V(x-x')$ at all possible grid points $(x,x')$.  This yields a matrix; $\rho(x')$ becomes a vector, and the Hartree potential becomes a simple matrix-vector product scaled by $\Delta x$.  With these potentials, we propagate (\ref{eqn:tdks1d}) using operating splitting---the propagation scheme is
\[
\bvarphi((k+1) \Delta t) = \exp(-i (\Delta t/2) (-1/2) D_2) \exp(-i \Delta t \mathbf{V}(k \Delta t) ) \exp(-i (\Delta t/2) (-1/2) D_2) \bvarphi(k \Delta t).
\]
The potential changes from one time step to the next and is defined by evaluating the following quantity at each spatial grid point $x = x_j$:
\[
\mathbf{V}(k \Delta t) = \operatorname{diag}\left[ v^\text{ext}(x) + \frac{1}{2}v^\text{H}[\boldsymbol{\rho}(k \Delta t)](x) + v^\text{C}(x, k \Delta t) \right]
\]
with $\brho(k \Delta t) = 2 | \bvarphi(k \Delta t) |^2$.  Propagating from $k=0$ to $k = N_T-1$, we obtain a $\bvarphi$ trajectory and hence a $\brho$ trajectory as well.  The mean $L_2$ error between this TDKS-computed density and the reference density computed from the wave function (\ref{eqn:exactpsi}) is
\[
\| \rho^\text{TDKS} - \rho^\text{exact} \| = \frac{1}{N^T} \sum_{k=1}^{N_T} \left[ \sum_{j} ( \rho^\text{TDKS}(j \Delta x, k \Delta t) - \rho^\text{exact}(j \Delta x, k \Delta t) )^2 \Delta x \right]^{1/2}.
\]
At the grid resolutions $N$ and $N_T$ given in Table 1 of the manuscript:
\[
(N, N_T) \in \left\{(301, 5 \times 10^3), (601, 5 \times 10^3), (1201, 5 \times 10^4), (2401, 5 \times 10^4), (4801, 2 \times 10^5) \right\},
\]
we find that this error is equal to
\[
8.08 \times 10^{-4}, 2.83 \times 10^{-4}, 3.83 \times 10^{-5}, 2.63 \times 10^{-5}, 6.58 \times 10^{-6},
\]
respectively.  Given all the numerical approximations required to produce both $\rho^\text{TDKS}$ and $\rho^\text{exact}$, this constitutes close quantitative agreement.  Additionally, we see that $\rho^\text{TDKS}$ is converging to $\rho^\text{exact}$ as grid sizes increase.  We conclude that exact inversion in 1D has been implemented correctly.

\subsection{TD-CI Reference States}

\subsubsection{Helium atom, Field $0.15$ in $z$ direction}

Selected CIS States, showing CI coefficients of magnitude 0.1 or higher. The first and second states are marked up to give a guide for reading the remaining cases.

\begin{itemize}
    \item \textbf{CI state 1} ($S^2 = 0.00000$, pure singlet)
    \begin{itemize}
        \item HF: $\phantom{-}0.99996$ (Hartree-Fock dominated state)
    \end{itemize}
    
    \item \textbf{CI state 2} ($S^2 = 0.00000$)
    \begin{itemize}
        \item $\beta(1 \rightarrow 2)$: $\phantom{-}0.21560$ (beta electron excited from HOMO to LUMO)
        \item $\beta(1 \rightarrow 4)$: $\phantom{-}0.66775$ (beta electron excited from HOMO to LUMO+2)
        \item $\alpha(1 \rightarrow 2)$: $\phantom{-}0.21560$ (alpha electron excited from HOMO to LUMO)
        \item $\alpha(1 \rightarrow 4)$: $\phantom{-}0.66775$ (alpha electron excited from HOMO to LUMO+2)
    \end{itemize}
    
    \item \textbf{CI state 3} ($S^2 = 0.00000$)
    \begin{itemize}
        \item $\beta(1 \rightarrow 12)$: $-0.18970$
        \item $\beta(1 \rightarrow 14)$: $-0.67767$
        \item $\alpha(1 \rightarrow 12)$: $-0.18970$
        \item $\alpha(1 \rightarrow 14)$: $-0.67767$
    \end{itemize}
\end{itemize}

\vspace{1em}
\noindent\textbf{TD-CI energies and superposition coefficients (all real)}
\begin{align*}
    \text{Eigenvalues:} & \quad -2.862426, -1.942175, -0.947354 \\
    \text{Eigen gaps:}  & \quad \phantom{-}0.0000000, \phantom{-}0.9202502, \phantom{-}1.9150718 \\
    \text{Initial } c:  & \quad -0.9996130, -0.0259265, \phantom{-}0.0100818
\end{align*}

\noindent\textbf{Time points selected for inversion (a.u.):} \\
0.0, 1.5, 3.0, 4.5, 6.0, 7.5, 9.0, 10.5, 12.0, 15.0

\noindent\textbf{Extrapolation times:} \\
15.6, 16.2, 16.8, 17.4, 18.0, 18.6, 19.2, 19.8, 20.4, 21.0

\subsubsection{Helium atom, Field $0.30$ in $z$ direction}

Selected CIS States, showing CI coefficients of magnitude 0.1 or higher.

\begin{itemize}
    \item \textbf{CI state 1} ($S^2 = 0.00000$)
    \begin{itemize}
        \item HF: $\phantom{-}0.99999$
    \end{itemize}
    
    \item \textbf{CI state 2} ($S^2 = 0.00000$)
    \begin{itemize}
        \item $\beta(1 \rightarrow 2)$: $\phantom{-}0.33658$
        \item $\beta(1 \rightarrow 4)$: $-0.60750$
        \item $\alpha(1 \rightarrow 2)$: $\phantom{-}0.33658$
        \item $\alpha(1 \rightarrow 4)$: $-0.60750$
    \end{itemize}
    
    \item \textbf{CI state 3} ($S^2 = 0.00000$)
    \begin{itemize}
        \item $\beta(1 \rightarrow 2)$: $-0.61958$
        \item $\beta(1 \rightarrow 4)$: $-0.32686$
        \item $\alpha(1 \rightarrow 2)$: $-0.61958$
        \item $\alpha(1 \rightarrow 4)$: $-0.32686$
    \end{itemize}
    
    \item \textbf{CI state 4} ($S^2 = 0.00000$)
    \begin{itemize}
        \item $\beta(1 \rightarrow 12)$: $\phantom{-}0.57419$
        \item $\beta(1 \rightarrow 13)$: $-0.39991$
        \item $\alpha(1 \rightarrow 12)$: $\phantom{-}0.57419$
        \item $\alpha(1 \rightarrow 13)$: $-0.39991$
    \end{itemize}
    
    \item \textbf{CI state 5} ($S^2 = 0.00000$)
    \begin{itemize}
        \item $\beta(1 \rightarrow 6)$: $\phantom{-}0.69926$
        \item $\alpha(1 \rightarrow 6)$: $\phantom{-}0.69926$
    \end{itemize}
\end{itemize}

\vspace{1em}
\noindent\textbf{TD-CI energies and superposition coefficients (all real)}
\begin{align*}
    \text{Eigenvalues:} & \quad -2.865101, -2.119230, -1.920792, -1.574094, -0.946481 \\
    \text{Eigen gaps:}  & \quad \phantom{-}0.0000000, \phantom{-}0.7458709, \phantom{-}0.9443092, \phantom{-}1.2910068, \phantom{-}1.9186199 \\
    \text{Initial } c:  & \quad \phantom{-}0.9982571, \phantom{-}0.0290904, \phantom{-}0.0469763, \phantom{-}0.0050468, -0.0201044
\end{align*}

\noindent\textbf{Time points selected for inversion (a.u.):} \\
0.0, 0.9, 1.8, 2.7, 3.6, 4.5, 5.4, 6.3, 7.2, 8.1, 9.0, 10.5, 12.0, 15.0, 18.0, 21.0, 24.0, 27.0

\noindent\textbf{Extrapolation times:} \\
27.6, 28.2, 28.8, 29.4, 30.0, 30.6, 31.2, 31.8, 32.4, 33.0

\subsubsection{Helium atom, FCI, Field $0.15$ in $z$ direction}

Selected FCI States, showing CI coefficients of magnitude 0.1 or higher. The first 12 orbitals are shown in the CI vectors.

\begin{itemize}
    \item \textbf{CI state 1} ($S^2 = 0.00000$)
    \begin{itemize}
        \item \texttt{100000000000} $\vert$ \texttt{100000000000} : $\phantom{-}0.99594$
    \end{itemize}
    
    \item \textbf{CI state 2} ($S^2 = 0.00000$)
    \begin{itemize}
        \item \texttt{100000000000} $\vert$ \texttt{010000000000} : $-0.21121$
        \item \texttt{100000000000} $\vert$ \texttt{000100000000} : $-0.65564$
        \item \texttt{010000000000} $\vert$ \texttt{100000000000} : $-0.21121$
        \item \texttt{000100000000} $\vert$ \texttt{100000000000} : $-0.65564$
    \end{itemize}
    
    \item \textbf{CI state 3} ($S^2 = 0.00000$)
    \begin{itemize}
        \item \texttt{000000000001} $\vert$ \texttt{100000000000} : $-0.18588$
        \item \texttt{000000000000} $\vert$ \texttt{100000000000} : $-0.66807$
        \item \texttt{100000000000} $\vert$ \texttt{000000000001} : $-0.18588$
        \item \texttt{100000000000} $\vert$ \texttt{000000000000} : $-0.66807$
    \end{itemize}
    
    \item \textbf{CI state 4} ($S^2 = 0.00000$)
    \begin{itemize}
        \item \texttt{100000000000} $\vert$ \texttt{010000000000} : $\phantom{-}0.65765$
        \item \texttt{100000000000} $\vert$ \texttt{000100000000} : $-0.20831$
        \item \texttt{010000000000} $\vert$ \texttt{100000000000} : $\phantom{-}0.65765$
        \item \texttt{010000000000} $\vert$ \texttt{010000000000} : $\phantom{-}0.10142$
        \item \texttt{000100000000} $\vert$ \texttt{100000000000} : $-0.20831$
    \end{itemize}
    
    \item \textbf{CI state 5} ($S^2 = 0.00000$)
    \begin{itemize}
        \item \texttt{000001000000} $\vert$ \texttt{000001000000} : $-0.16734$
        \item \texttt{000001000000} $\vert$ \texttt{100000000000} : $\phantom{-}0.68970$
        \item \texttt{100000000000} $\vert$ \texttt{000001000000} : $\phantom{-}0.68970$
    \end{itemize}
\end{itemize}

\vspace{1em}
\noindent\textbf{TD-CI energies and superposition coefficients (all real)}
\begin{align*}
    \text{Eigenvalues:} & \quad -2.903539, -2.149002, -2.009312, -1.644640, -1.009112 \\
    \text{Eigen gaps:}  & \quad \phantom{-}0.0000000, \phantom{-}0.7545367, \phantom{-}0.8942272, \phantom{-}1.2588992, \phantom{-}1.8944271 \\
    \text{Initial } c:  & \quad \phantom{-}0.9995605, \phantom{-}0.0093010, -0.0264280, \phantom{-}0.0011225, \phantom{-}0.0096228
\end{align*}

\noindent\textbf{Time points selected for inversion (a.u.):} \\
0.0, 1.5, 3.0, 4.5, 6.0, 7.5, 9.0, 10.5, 12.0, 15.0

\noindent\textbf{Extrapolation times:} \\
15.6, 16.2, 16.8, 17.4, 18.0, 18.6, 19.2, 19.8, 20.4, 21.0

\subsubsection{Dihydrogen molecule, Field $0.15$ in $z$ direction}

Selected CIS States, showing CI coefficients of magnitude 0.1 or higher.

\begin{itemize}
    \item \textbf{CI state 1}
    \begin{itemize}
        \item HF: $\phantom{-}0.99950$
    \end{itemize}
    
    \item \textbf{CI state 2}
    \begin{itemize}
        \item $\beta(1 \rightarrow 2)$: $\phantom{-}0.64118$
        \item $\beta(1 \rightarrow 3)$: $\phantom{-}0.29358$
        \item $\alpha(1 \rightarrow 2)$: $\phantom{-}0.64118$
        \item $\alpha(1 \rightarrow 3)$: $\phantom{-}0.29358$
    \end{itemize}
    
    \item \textbf{CI state 3}
    \begin{itemize}
        \item $\beta(1 \rightarrow 2)$: $\phantom{-}0.28436$
        \item $\beta(1 \rightarrow 3)$: $-0.63566$
        \item $\beta(1 \rightarrow 6)$: $\phantom{-}0.11915$
        \item $\alpha(1 \rightarrow 2)$: $\phantom{-}0.28436$
        \item $\alpha(1 \rightarrow 3)$: $-0.63566$
        \item $\alpha(1 \rightarrow 6)$: $\phantom{-}0.11915$
    \end{itemize}
\end{itemize}

\vspace{1em}
\begin{align*}
    \text{Eigenvalues:} & \quad -1.137734, -0.661415, -0.522782 \\
    \text{Eigen gaps:}  & \quad \phantom{-}0.0000000, \phantom{-}0.4763189, \phantom{-}0.6149513 \\
    \text{Initial } c:  & \quad -0.9965955, -0.0753436, -0.0334781
\end{align*}

\noindent\textbf{Time points selected for inversion (a.u.):} \\
0.0, 0.3, 0.6, 0.9, 1.2, 1.5, 3.0, 4.5, 6.0, 7.5, 9.0, 10.5, 12.0, 13.5, 15.0, 16.5, 18.0, 19.5, 21.0, 22.5, 24.0, 25.5

\noindent\textbf{Extrapolation times:} \\
26.1, 26.7, 27.3, 27.9, 28.5, 29.1, 29.7, 30.3, 30.9, 31.5

\subsubsection{Dihydrogen molecule, Field $0.15$ in $z$ direction (Second Set)}

Selected CIS States, showing CI coefficients of magnitude 0.1 or higher.

\begin{itemize}
    \item \textbf{CI state 1}
    \begin{itemize}
        \item HF: $-0.99999$
    \end{itemize}
    
    \item \textbf{CI state 2}
    \begin{itemize}
        \item $\beta(1 \rightarrow 2)$: $-0.64134$
        \item $\beta(1 \rightarrow 3)$: $-0.29335$
        \item $\alpha(1 \rightarrow 2)$: $-0.64134$
        \item $\alpha(1 \rightarrow 3)$: $-0.29335$
    \end{itemize}
    
    \item \textbf{CI state 3}
    \begin{itemize}
        \item $\beta(1 \rightarrow 2)$: $-0.28443$
        \item $\beta(1 \rightarrow 3)$: $\phantom{-}0.63576$
        \item $\beta(1 \rightarrow 6)$: $-0.11842$
        \item $\alpha(1 \rightarrow 2)$: $-0.28443$
        \item $\alpha(1 \rightarrow 3)$: $\phantom{-}0.63576$
        \item $\alpha(1 \rightarrow 6)$: $-0.11842$
    \end{itemize}
    
    \item \textbf{CI state 4}
    \begin{itemize}
        \item $\beta(1 \rightarrow 6)$: $\phantom{-}0.69300$
        \item $\alpha(1 \rightarrow 6)$: $\phantom{-}0.69300$
    \end{itemize}
    
    \item \textbf{CI state 5}
    \begin{itemize}
        \item $\beta(1 \rightarrow 7)$: $\phantom{-}0.69737$
        \item $\alpha(1 \rightarrow 7)$: $\phantom{-}0.69737$
    \end{itemize}
\end{itemize}

\vspace{1em}
\begin{align*}
    \text{Eigenvalues:} & \quad -1.137734, -0.661415, -0.522782, -0.287345, -0.058092 \\
    \text{Eigen gaps:}  & \quad \phantom{-}0.0000000, \phantom{-}0.4763189, \phantom{-}0.6149513, \phantom{-}0.8503883, \phantom{-}1.0796419 \\
    \text{Initial } c:  & \quad \phantom{-}0.9961074, -0.0753067, \phantom{-}0.0334617, -0.0308694, -0.0051405
\end{align*}

\noindent\textbf{Time points selected for inversion (a.u.):} \\
0.0, 0.3, 0.6, 0.9, 1.2, 1.5, 3.0, 4.5, 6.0, 7.5, 9.0, 10.5, 12.0, 13.5, 15.0, 16.5, 18.0, 19.5, 21.0, 22.5, 24.0, 25.5

\noindent\textbf{Extrapolation times:} \\
26.1, 26.7, 27.3, 27.9, 28.5, 29.1, 29.7, 30.3, 30.9, 31.5

\subsection{Slater Basis Sets}

The following Slater basis sets were used for inverse calculations to obtain $v_{XC}(r,t)$. Values are exponential coefficients. All $l > 0$ basis functions use real spherical harmonics. Integrals were evaluated using the SlaterGPU library.

\subsubsection{Helium}
\begin{description}
    \item[$1s$:] 8.000, 5.333, 3.556, 2.370, 1.580, 1.053, 0.702, 0.468, 0.312, 0.208, 0.139
    \item[$2p$:] 5.243, 3.277, 2.048, 1.280, 0.800, 0.500, 0.333, 0.222, 0.148
    \item[$3d$:] 3.0, 1.8
\end{description}

\subsubsection{Hydrogen}
\begin{description}
    \item[$1s$:] 3.00, 2.07, 1.45, 1.00, 0.68, 0.46
    \item[$2p$:] 3.50, 2.30, 1.50, 1.00, 0.60
    \item[$3d$:] 2.00
\end{description}


%

\end{document}